\documentclass[final]{l4dc2024}

\title[Attentive Recurrent Neural Cellular Automata]{Learning Locally Interacting Discrete Dynamical Systems: Towards Data-Efficient and Scalable Prediction}
\usepackage{times}
\coltauthor{\Name{Beomseok Kang} \Email{beomseok@gatech.edu}\\
 \Name{Harshit Kumar} \Email{hkumar64@gatech.edu}\\
 \Name{Minah Lee} \Email{minah.lee@gatech.edu}\\
 \Name{Biswadeep Chakraborty} \Email{biswadeep@gatech.edu}\\
 \Name{Saibal Mukhopadhyay} \Email{smukhopadhyay6@gatech.edu}\\
 \addr School of Electrical and Computer Engineering, Georgia Institute of Technology, GA, USA}

\usepackage{times}
\usepackage{array}
\usepackage{multirow}
\usepackage{hhline}
\usepackage{wrapfig,lipsum,booktabs}
\usepackage{bbm}
 
\begin{document}

\maketitle

\begin{abstract}%
Locally interacting dynamical systems, such as epidemic spread, rumor propagation through crowd, and forest fire, exhibit complex global dynamics originated from local, relatively simple, and often stochastic interactions between dynamic elements. Their temporal evolution is often driven by transitions between a finite number of discrete states. Despite significant advancements in predictive modeling through deep learning, such interactions among many elements have rarely explored as a specific domain for predictive modeling. We present \textbf{A}ttentive \textbf{R}ecurrent \textbf{N}eural \textbf{C}ellular \textbf{A}utomata (AR-NCA), to effectively discover unknown local state transition rules by associating the temporal information between neighboring cells in a permutation-invariant manner. AR-NCA exhibits the superior generalizability across various system configurations (\textit{i.e.}, spatial distribution of states), data efficiency and robustness in extremely data-limited scenarios even in the presence of stochastic interactions, and scalability through spatial dimension-independent prediction. Our code and supplementary material are available in \url{https://github.com/beomseokg/ARNCA}.
\end{abstract}


\section{Introduction}
In natural systems, seemingly simple local interactions among dynamic elements give rise to complex global behaviors. Consider, for instance, the spread of epidemics \citep{ghosh2020data, white2007modeling}, the propagation of rumors in social networks \citep{wang2014social, kawachi2008rumor}, the dynamics governing forest fires \citep{karafyllidis1997model, trunfio2004predicting}. Their temporal evolution is often driven by transitions between a finite number of discrete states (\textit{e.g.}, healthy or infected in the epidemics). Learning such systems poses interesting questions emerged from the localized dynamics. First, prediction models may not require many training samples since few sequential observation during the system evolution involves frequent state transitions. Also, the prediction in unobserved larger systems may not be degraded since the systems can be considered duplicates of smaller systems. While data efficient and scalable learning are challenging problems for deep learning-based predictive modeling, our hypothesis is that such challenges can be effectively resolved if models essentially learn state transition rules. However, how to discover the unknown rules governing discrete-state dynamical systems, particularly those influenced by the local interactions, remains relatively unexplored as specific deep learning problems.

Deep learning-based predictive modeling has shown substantial advancements in recent years, especially for learning trajectories in continuous state space. For examples, network dynamics \citep{zang2020neural}, scene/video \citep{wang2017predrnn}, and multi-agent movement \citep{kang2023forecasting} predictions are related to our problem in the sense of having many dynamic elements (\textit{e.g.}, nodes or pixels) in two-dimensional spatiotemporal systems. However, they do not explicitly capture the local interactions, demonstrating specious prediction results with poor generalizability in the above-mentioned scenarios. Conventionally, cellular automata with human-designed state transition rules have long served as an effective computational function for simulating discrete dynamical systems with the local interactions \citep{bauer2009agent, macal2009agent, clarke2014cellular}. Recent progress in cellular automata has embraced deep learning to replace the state transition rules with learnable non-linear functions, often referred to as neural cellular automata (NCA) \citep{grattarola2021learning, tesfaldet2022attention, mordvintsev2020growing}. Nonetheless, these applications have predominantly focused on static and deterministic environments such as image processing, where temporal information about current cells is not essential to predict future states. 

In this paper, we introduce \textbf{A}ttentive \textbf{R}ecurrent \textbf{N}eural \textbf{C}ellular \textbf{A}utomata (AR-NCA), which is a novel NCA architecture specifically designed for learning locally interacting discrete dynamical systems. AR-NCA involves a recurrent cellular attention module that couples long short-term memory (LSTM) \textcolor{blue}{\citep{hochreiter1997long}} and cellular self-attention. Cellular self-attention module associates the temporal information of each cell with its local neighborhood in a permutation-invariant manner. It promotes the center cell and its neighboring cells from any direction to effectively contribute to the discovery of their unknown interaction rules. Our technical contribution is primarily in identifying the interesting properties emerged from this novel NCA-based architecture. For example, the permutation invariance in AR-NCA leads to efficient discovery of the hidden interaction rules in extremely data-limited scenarios even in the presence of stochastic interactions (\textbf{Data efficiency}). Furthermore, the cell-based processing of AR-NCA allows to train in relatively small systems and then apply to 256$\times$ larger systems without re-training and degradation of the performance (\textbf{Scalability}). AR-NCA is evaluated in the task to predict the evolution of a certain state within the three synthetic systems; forest fire \citep{tisue2004netlogo}, host-pathogen \citep{sayama2013pycx}, and stock market \citep{wei2003cellular} models.

\section{Background and Dataset}
\paragraph{System Formulation} Let us consider a system in the $n\times n$ two-dimensional space ($\mathcal{R}^{n \times n}$) representing $n^{2}$ cells, where each cell is spatially stationary and interacting with its neighboring cells. The system includes a discrete state variable ($S$) that is evolved by an unknown function (\(g\)). For example, the state of each cell at the $t$-th timestep, $i$-th row, and $j$-th column, is defined by \(S_{(i,j)}(t) \in \{s_{1}, s_{2}, \cdots, s_{m}\} \), where \(s_{k}\) is the $k$-th state in $m$-degree state space. Then, the state is changed over time by \(S_{(i,j)}(t+1)=g(S_{N'_{(i,j)}}(t), S_{N'_{(i,j)}}(t-1), ...)\), where \(S_{N'_{(i,j)}}\) is the set of states in 3$\times$3 cells including the centered one at (\(i,j\)) and its neighboring ones \(N_{(i,j)}\). 

\paragraph{Problem Statement} We are particularly interested in a certain state, saying $s^{*}$. In other words, our ground truth is a \(n \times n\) binary grid that represents whether the state of each cell is $s^{*}$ or not (\textit{i.e.}, \(\mathbbm{1}_{S_{i,j}(t)=s^{*}}\)). We assume that prediction models observe the first $t_{\text{obs}}$ timesteps from $t=0$ and predict the next $t_{\text{pred}}-t_{\text{obs}}$ timesteps. Then, our objective is as follows:
\vspace{-2mm}
\begin{equation}
\min \sum_{t=t_{\text{obs}}}^{t_{\text{pred}}-1}\sum_{i,j}{E(\mathbbm{1}_{S_{(i,j)}(t)=s^{*}}, f_{\theta}(i,j,t,S(t=0), S(t=1), ... , S(t=t_{\text{obs}-1})))}
\label{equation:objective}
\end{equation}
\noindent where $f_{\theta}$ is a neural network-based function parameterized by $\theta$, and $E$ is an error metric. The function needs to identify the unknown \(g\) from training data, possibly by a data-efficient and scalable manner. We investigate how to learn the interactions ($g$) and minimize (\ref{equation:objective}) from data using $f_{\theta}$.
\vspace{-2mm}
\begin{figure}[t]
\centering
\includegraphics[width=0.9\columnwidth]{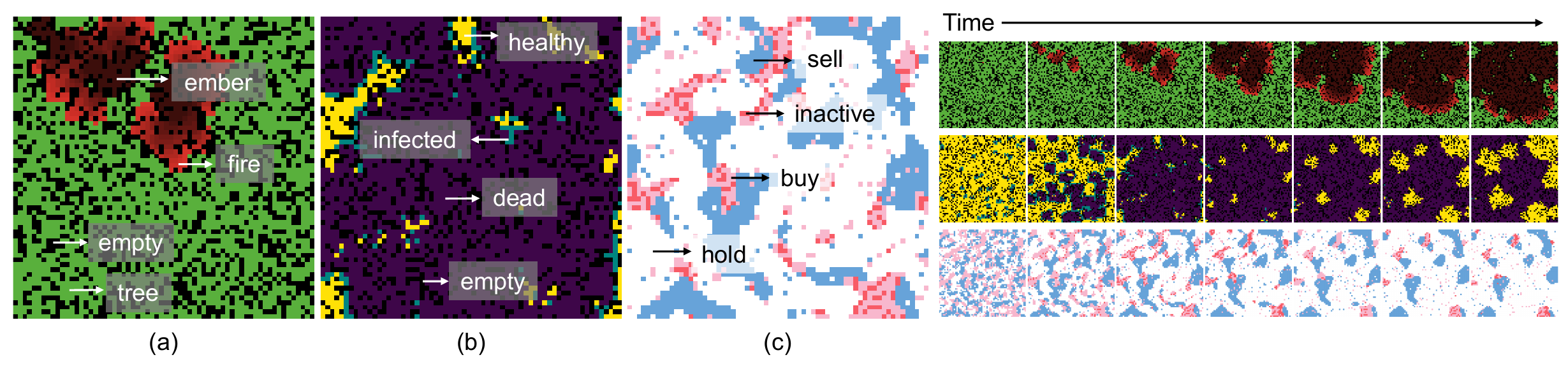}
\vspace*{-5mm}
\caption{Synthetic discrete dynamical systems (a) forest fire model, (b) host-pathogen model, (c) stock market model. Their states are represented by distinct colors.}
\label{figure_dataset}
\vspace*{-9mm}
\end{figure}
%\vspace*{-9mm}
\paragraph{Dataset} We consider the three synthetic environments, forest fire, host-pathogen, and stock market models, as examples of the above-mentioned dynamical system and our learning environments\footnote{More details are available in the supplementary material.}. Each system includes four discrete states as described in Figure \ref{figure_dataset}, but driven by different temporal dependency. For example, in the host-pathogen model, the next state is simply depending on the current state. In contrast, we introduce an inactive state in the stock market model, which depends on two timesteps (current and past). The forest fire model has another internal states (heat values, $q(t)$), which can depend on more than two timesteps. Our implementation are based on the prior works \citep{tisue2004netlogo, sayama2013pycx, wei2003cellular}.
\vspace*{-1.5mm}
\paragraph{Forest Fire model} It is a well-known example of self-organizing complex system. We modify the interaction rule to be non-trivial based on the prior works \citep{rothermel1972mathematical, chen1990deterministic}. The model involves empty, tree, fire, and ember cells. Our $s^{*}$ is the burning state (ember and fire). It evolves as follows: \textbf{1)} Create 64$\times$64 grid filled with randomly distributed tree and empty cells. \textbf{2)} Define fire seeds with \(q_{\text{seed}}=6\) (initial heat) at three random locations. \textbf{3)} Each tree cell accumulates the heat from neighboring fire and ember cells by \(q_{(i,j)}(t+1)=q_{(i,j)}(t) + q_{\text{transfer}} \sum_{n \in \text{Neighbor}}{q_{n}(t)}\), \(q_{\text{transfer}}=0.3\). \textbf{4)} Check if the current heat value exceeds \(q_{\text{threshold}}=3\) (If so, tree$\rightarrow$fire). \textbf{5)} Decrease the heat values of fire and ember cells by \(q(t+1)=q(t) - q_{\text{die}}\), \(q_{\text{die}}=1\). \textbf{6)} For
fire cells, transitioned to ember cells. Repeat from \textbf{3)} to \textbf{6)}. In stochastic settings, we consider the probability \(p_{\text{heat}}=0.9\) to randomly ignore the neighboring cell (\textit{i.e.}, \(q_{n}(t)=0\)) in \textbf{3)}.
\vspace*{-1.5mm}
\paragraph{Host-Pathogen model} It is originated from the theoretical understanding of the interaction between viruses and host organisms in a population level \citep{sayama2013pycx}. The model includes empty, dead, healthy, and infected cells. Our $s^{*}$ is the healthy state. It evolves as follows: \textbf{1)} Create 64$\times$64 grid filled with randomly distributed infected (1\%), healthy (75\%), and empty (the rest) cells. \textbf{2)} For dead cells, cured by each of neighboring healthy cells with the probability \(p_{\text{cure}}=0.15\). \textbf{4)} For healthy cells, infected by each of neighboring infected cells with probability \(p_{\text{infect}}=0.85\). \textbf{5)} For infected cells, transitioned to dead cells. Repeat from \textbf{2)} to \textbf{5)}. 
\vspace*{-1.5mm}
\paragraph{Stock Market model} The complexity of investor's behavior in the stock market has been studied through cellular automata \citep{wei2003cellular}. They aimed to model the imitation behavior of investors affected by neighboring investors. The model has hold, sell, buy, and (additional) inactive cells. Our $s^{*}$ is the buying state. It evolves as follows: \textbf{1)} Create 64$\times$64 grid filled with randomly chosen states (\textit{e.g.}, buy, sell, hold). \textbf{2)} Each cell makes the stochastic state transition based on neighboring cells. \textbf{3)} The cells which have been in two consecutive buying states become inactive. \textbf{4)} The inactive cells in the previous sequence are transitioned to the buying state. Repeat from \textbf{2)} to \textbf{4)}. There is a transition matrix \citep{wei2003cellular} parameterized by \(M=0.05\) (positive market status) and \(p_{\text{invest}}=0.95\) to define the stochastic state transition. The dominant state of neighboring cells determines the next state of the center cell.

\vspace{-3mm}
\section{Related Work on Learning Dynamical Systems}

Prior studies on interaction learning have been performed in the context of multi-agent dynamical systems. However, they mostly consider few agents (\(\sim\)10) using a complete graph or pre-defined interaction graph \citep{sankararaman2019social, liu2020multi, li2021grin, niu2021multi}. Other works, more related to our problem, investigated complex physical systems with many nodes, such as particle systems \citep{sanchez2020learning} or heat diffusion \citep{zang2020neural} in a grid, using graph networks. Nonetheless, their focus lies in modeling continuous trajectories in state space (\textit{e.g.}, position of particle), not classifying discrete states. In addition, they often exhibit high computational costs, quadratically (or higher) increasing to the number of dynamic elements \citep{zang2020neural, battaglia2016interaction}. Video prediction networks have been widely employed in various sequential modeling problems, such as physical systems \citep{finn2016unsupervised}, driving scenes \citep{xu2017end}, and weather forecasting \citep{shi2015convolutional}. These approaches are able to efficiently process numerous dynamic elements, represented by pixels, without any assumption or knowledge about the systems. Hence, we primarily compare AR-NCA with several video prediction networks.

\paragraph{Neural Cellular Automata} The primary mechanism of NCA is in its Moore's neighborhood (\textit{i.e.}, 3$\times$3 cells)-based operation that gradually updates the center cell through iterative inferences. This effectively models local interactions existing in the systems. In this light, the popular application of NCA has been image denoising \citep{tesfaldet2022attention} and texture generation \citep{pajouheshgar2023dynca, mordvintsev2020growing}. However, prior studies mostly employed convolutional neural network (CNN)-based architectures, challenging to learn dynamical systems due to the following two reasons. First, the NCAs without memory functions hinders modeling the dynamic cells if the transition rule of cells is depending on their temporal information. Second, iterative inferences for sequential data create a long computational graph, making both training and inference unstable and computationally expensive.

\vspace{-3mm}
\section{Proposed Approach}

AR-NCA is specifically designed for the dynamical systems to discover unknown local interactions. Before the architectural details, we first discuss the motivation behind the design of AR-NCA.
\vspace*{-1.5mm}
\paragraph{Lack of memory function in existing NCAs}
Memory is an important function to model the evolution of the dynamical systems since the next cell state can depend on not only current but also past states. For example, in forest fire model, even if cells are in the same ember states, newer ember cells ($q \neq 0$) can still radiate heats to neighboring cells while older ones ($q = 0$) do not. In this case, the time when the fire state is transitioned to the ember is important information to store in the prediction models. However,  most prior NCA networks are based on CNN \citep{mordvintsev2020growing, pajouheshgar2023dynca, gilpin2019cellular}. Recently, graph networks \citep{grattarola2021learning}, transformer \citep{tesfaldet2022attention}, and variational autoencoder \citep{palm2022variational} are coupled within NCAs, which do not explicitly have memory functions, such as recurrent networks.
\begin{figure}[t]
\centering
\includegraphics[width=0.9\columnwidth]{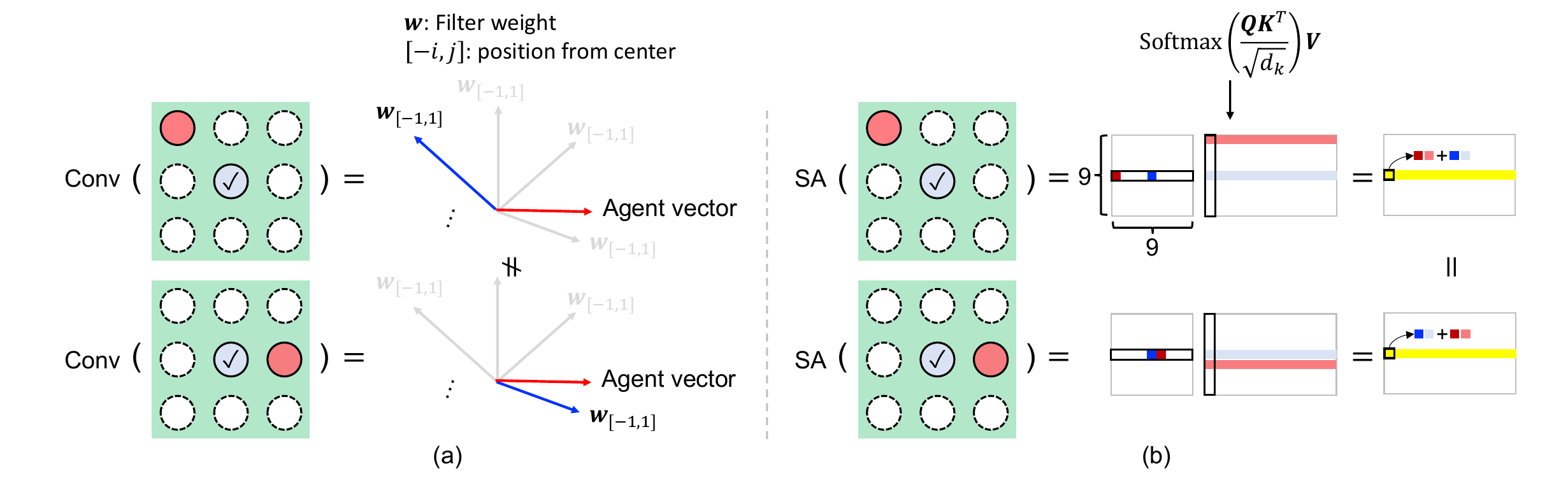}
\vspace*{-5mm}
\caption{Spatial dependency on cell location in (a) convolution (Conv) and (b) self-attention (SA).}
\label{figure_attentionandconv}
\vspace{-7mm}
\end{figure}

\vspace*{-1.5mm}
\paragraph{Spatial dependency in CNN-based NCAs} While CNN-based architectures are prevalent in NCAs, another challenge is the spatial dependency in estimating local interactioncs. Let us consider two configurations of 3$\times$3 cells as shown in Figure \ref{figure_attentionandconv}, assuming the red and blue cells are interacting (white cell is empty). Figure \ref{figure_attentionandconv}(a) describes the convolution operation on the red cell as an inner product between the weight vector (blue) and the encoded vector of the red cell (red). However, the weight vector is not identical across the different cell locations, indicating that the convolution output depends on the spatial order of neighboring cells\footnote{Empirical results on the unbalanced interaction learning of CNN are provided in the supplementary material.}. In order to measure the interaction regardless of the red cell's location, all the nine weight vectors should be properly shifted together. This requires the model to be exposed by a various set of spatial configurations, possibly by more training samples. In contrast, Figure \ref{figure_attentionandconv}(b) shows that self-attention operation estimates the interaction through attention scores (9$\times$9 matrix), where the red cell's location is no longer important in the final yellow vector (\textit{i.e.}, permutation invariance). 

\vspace{-3mm}
\subsection{Attentive Recurrent Neural Cellular Automata}

Cellular automata (CA) is generally designed with the cell states and update rules as a function of the states \citep{mordvintsev2020growing}. We aim to design a neural network-based non-trivial function (\textit{i.e.}, update rules) operating on the continuous vectors that represent the states of cells. Mathematically, our design space is primarily defined on \(x_{(i,j)}(t+1)=f_{\theta; \mathrm{CA}}(x_{(i,j)}(t), x_{N_{(i,j)}(1)}, ... , x_{N_{(i,j)}(8)}(t)) \in \mathcal{R}^{u} \), where \(x_{(i,j)}(t)\) is the high-dimensional representation of a cell state at the location \((i,j)\) and time \(t\), and \(N_{(i,j)}\) is a set of eight neighboring cells' locations around \((i,j)\). It should be permutation invariant as following:
\vspace{-2mm}
\begin{equation}
\begin{split}
f_{\theta; \mathrm{CA}}(x_{(i,j)}(t), x_{{N_{(i,j)}(1)}}(t), x_{{N_{(i,j)}(2)}}(t), ... , , x_{{N_{(i,j)}(8)}}(t))= \\
f_{\theta; \mathrm{CA}}(x_{(i,j)}(t), x_{{N_{(i,j)}(\pi(1))}}(t), x_{{N_{(i,j)}(\pi(2))}}(t), ... , , x_{{N_{(i,j)}(\pi(8))}}(t))
\end{split}
\label{equation_permuationinvariance}
\vspace{-2mm}
\end{equation}
where \(\pi\) is every permutation of \(\{1, 2, ... , 8\}\). In addition to \(f_{\theta; \mathrm{CA}}\), we need two more functions, \(f_{\theta;\mathrm{Enc}}\) and \(f_{\theta;\mathrm{Dec}}\). The cells in our systems are observed in RGB space (\textit{i.e.}, $m=3$). Also, the ultimate task is to make the probabilistic estimation of a cell's certain state in the future. Hence, we need an encoding function that transfers the RGB vectors into the high-dimensional vectors and a decoding function that converts the high-dimensional vectors into probabilities. In summary, AR-NCA is composed of the three functions; cell-wise feature extractor (\(f_{\theta ;\mathrm{Enc}}:\mathcal{R}^{3} \rightarrow \mathcal{R}^{u}\)), recurrent cellular attention module (\(f_{\theta; \mathrm{CA}}:\mathcal{R}^{u} \rightarrow \mathcal{R}^{u}\)), and cell-to-probability decoder (\(f_{\theta;\mathrm{Dec}}:\mathcal{R}^{u} \rightarrow \mathcal{R}^{1}\)). 

\begin{figure}[t]
\centering
\includegraphics[width=0.9\columnwidth]{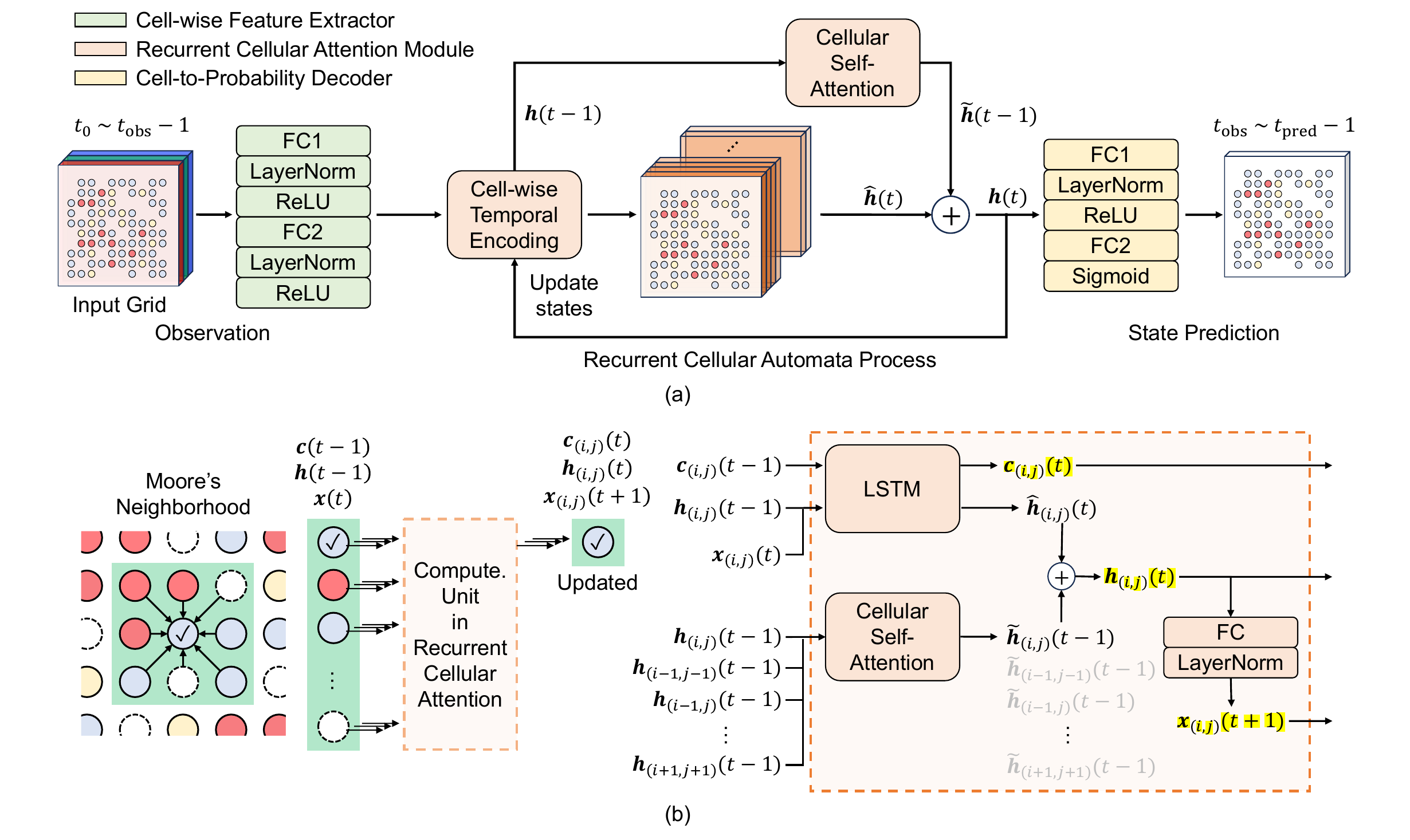}
\vspace{-5mm}
\caption{The overview of Attentive Recurrent Neural Cellular Automata architecture.}
\label{figure_networkarchitecture}
\vspace{-5mm}
\end{figure}
Figure \ref{figure_networkarchitecture}(a) shows the overview of the proposed network architecture. AR-NCA observes \(t_{{\text{obs}}}\) successive scenes of the system and predicts the next \(t_{{\text{pred}}}-t_{{\text{obs}}}\) scenes as probability maps. The cell-based feature extractor processes individual cell states represented by RGB colors during the observation period. The recurrent cellular attention module encodes the observed sequence of individual cell states, associates the encoded temporal information with neighboring cells by cellular self-attention, and predicts the next sequence in an autoregressive manner. The cell-to-probability decoder converts the encoded and predicted cell states into probabilities and creates probability maps. The feature extractor consists of two fully-connected (FC) layers. It is shared across entire cells since there is no spatial correlation in cell states. Note, the spatial dimensions of the systems do not change in the encoded space, and only the channel dimension varies. It is important to preserve the spatial dimensions because the interaction is very localized, indicating the individual cell's behavior is critical in modeling the interaction rules. The decoder is also designed with two fully-connected layers and shared for all the cells. As the final output should be a probability, sigmoid function is used as the last activation. Ground truth for the prediction is given by binary maps. The network is trained by binary cross-entropy (BCE) as an error metric ($E$) in (\ref{equation:objective}).

The \textbf{key feature of AR-NCA} is in the recurrent cellular attention module, which is the distinct module of AR-NCA. The motivation is that, the dynamics of each cell is updated to mimic the dynamics of neighboring cells. Considering the propagation of forest fire, spread of epidemics or rumors, the dynamics of each cell is totally altered after the interaction, and the new dynamics is analogous to that of its neighbor that induced the interaction. In this light, recurrent cellular attention module aims to efficiently deliver the dynamics of neighboring cells into the center cell. 

In order to deal with the cell-wise dynamics, we assume that the temporal information of each cell should be well-preserved in the module. For the reason, the cell-based temporal encoding is performed by a single-layer LSTM which is shared across the entire cells while having the independent hidden states for each of cells\footnote{We investigate the basic RNN-based AR-NCA as well. The results are available in the supplementary material.}. Now, the dynamics of each cell is translated into its hidden states in LSTM. Cellular self-attention aims to quickly update the hidden states of each cell by delivering the dynamics of neighboring cells. Figure \ref{figure_networkarchitecture}(b) describes the recurrent cellular attention mechanism in Moore's neighborhood. Formally, the hidden state (\(h_{(i,j)}(t)\)) for a cell at \((i,j)\) position at time \(t\) in recurrent cellular attention is given by:
\vspace*{-3mm}
\begin{equation}
h_{(i,j)}(t) = \tilde{h}_{(i,j)}(t-1) + \hat{h}_{(i,j)}(t) 
\label{equation_sum_of_hidden_states}
\vspace{-3mm}
\end{equation}
\noindent where \(\tilde{h}_{(i,j)}(t-1)\) is the output from the cellular self-attention module, representing the attention vector for the cell at \((i,j)\) associated with its' Moore's neighborhood (\textit{i.e.}, \(N_{(i,j)})\), and \(\hat{h}_{(i,j)}(t)\) is the hidden state modeled from the LSTM cell for the cell at \((i,j)\) at time \(t\). An intuitive understanding for (\ref{equation_sum_of_hidden_states}) is that, by summing the \((i,j)\) cell's hidden state and its neighbor's one, the dynamics of \((i,j)\) cell is quickly shifted to mimic the behavior of its neighbor. \(\tilde{h}_{(i,j)}(t-1)\) and \(\hat{h}_{(i,j)}(t)\) are given by:
\vspace*{-3mm}
\begin{equation}
\hat{h}_{(i,j)}(t)=o_{(i,j)}(t) \odot \text{tanh}(c_{(i,j)}(t))
\end{equation}
\vspace*{-5mm}
\begin{equation}
\tilde{h}_{(i,j)}(t-1)=\text{FC}(\text{softmax}(Q_{(i,j)} K_{(i,j)}^{\text{T}}) V_{(i,j)})_{(i,j)}
\end{equation}
\vspace*{-5mm}
\begin{equation}
Q_{(i,j)} = W_{\text{Q}} h_{N'_{(i,j)}}(t-1); \; K_{(i,j)} = W_{\text{K}} h_{N'_{(i,j)}}(t-1); \; V_{(i,j)} = W_{\text{V}} h_{N'_{(i,j)}}(t-1)
\end{equation}
\noindent where \(o_{(i,j)}\) and \(c_{(i,j)}\) are internal and cell states in the LSTM; $Q$, $K$, and $V$ are the query, key, and value in the self-attention module obtained by associated weight matrices $W_{Q}$, $W_{K}$, and $W_{V}$; \(h_{N'_{(i,j)}}\) is the set of hidden states representing 3$\times$3 cells located at (\(i,j\)) and \(N_{(i,j)}\). $\odot$ represents element-wise multiplication. In summary, recurrent cellular attention module couples the LSTM and cellular self-attention module to estimate the interaction from neighbor cells using the current hidden states and changes the future dynamics of the center cell to resemble the dynamics of important (\textit{i.e.}, high attention score) neighboring cells.

\vspace*{-3mm}
\section{Empirical Result}

\paragraph{Experimental Setting} The training is generally performed in 64$\times$64 scales with RGB channels. The three datasets include 700 chunks (train) and 300 chunks (test). Each chunk has 60 frames (forest fire) or 30 frames (others). Note, \(t_{\text{obs}}\) is 10 frames in common, and \(t_{\text{pred}}\) is 60 frames (forest fire) or 30 frames (others). We use F1 score and area under ROC curve (AUC) as evaluation metrics. Our F1 score is based on the probability threshold 0.5, and AUC assumes that we can have the different threshold for each prediction timestep. Both metrics are only measured in the prediction window (\(t_{\text{pred}}-t_{\text{obs}}\)) and averaged by the number of predicted frames. We set 300\(\sim\)1000 epochs, \(1e^{-4}\sim5e^{-4}\) learning rates, batch size 4 chunks and use Adam optimizer.

\begin{figure}
\centering
\includegraphics[width=0.9\columnwidth]{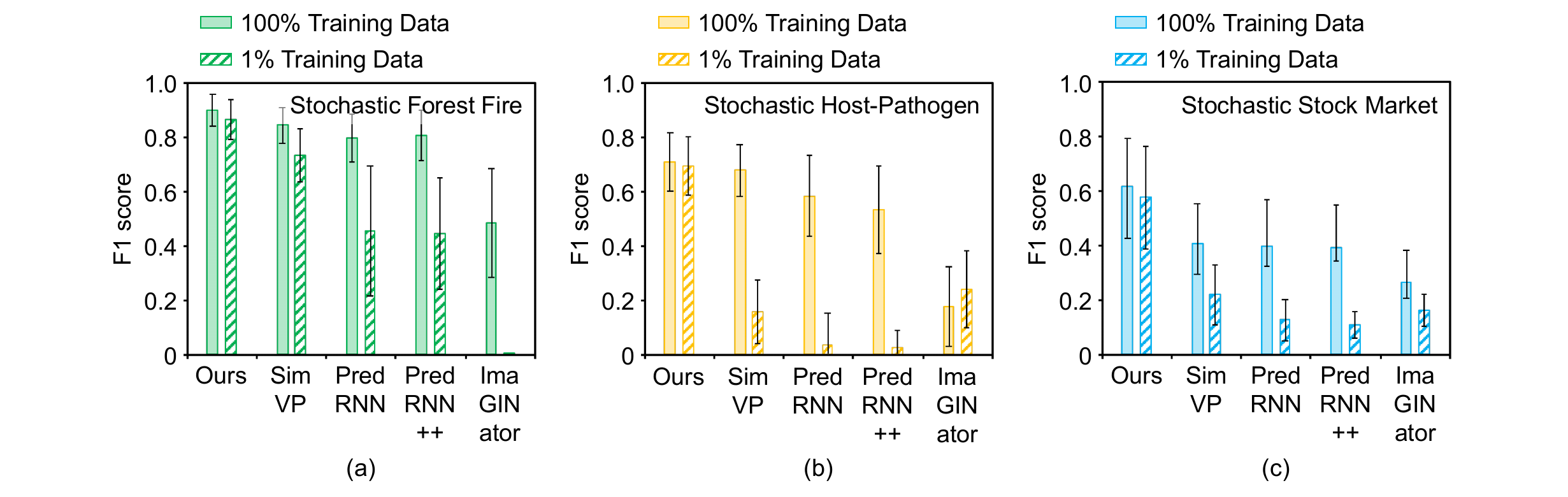}
\vspace{-5mm}
\caption{Comparison of data efficiency with video prediction networks in stochastic (a) forest fire, (b) host-pathogen, and (c) stock market models. The bar and line indicate the mean and standard deviation.}
\vspace{-5mm}
\label{figure_dataefficiencyvideoprediction}
\end{figure}

\vspace*{-1.5mm}
\paragraph{Comparison with Video Prediction Networks} As shown in Figure \ref{figure_dataset}, the input can be thought of as images and hence, we can consider the problem as a video prediction problem, where modelling the pixel-level dynamics is crucial. We perform a comprehensive comparison between AR-NCA and video prediction networks, including RNN \citep{wang2017predrnn, wang2018predrnn++}, GAN \citep{wang2020imaginator}, and CNN-based architectures \citep{gao2022simvp}. The compared models are trained to predict RGB images and then segment the predicted images into binary grids by comparing the color norm with the ground-truth colors of cell states. Figure \ref{figure_dataefficiencyvideoprediction} showcases that the performance disparity between AR-NCA and the compared models is noticeable particularly in the data-limited scenarios. Also, in Table \ref{table_videoprediction}, we observe that the performance of the video prediction networks are highly varying depending on forest configurations while AR-NCA outperforms them with the less variation. As the interaction rules are identical regardless of the tree distribution, such high variation of the performance should not be observed if the prediction networks have successfully discovered state transition rules. In summary, AR-NCA provides another perspective to process this type of sequential images (governed by pixel-level dynamics) by treating each of pixels as locally interacting cells.

\begin{table}[t]
\renewcommand{\arraystretch}{1.3}
\setlength{\tabcolsep}{3pt}
\fontsize{8.5pt}{8.5pt}\selectfont
\centering
\begin{tabular}{lcccccc} 

\hline

Method & Dense Forest (F1\(\uparrow\)) & Sparse Forest (F1\(\uparrow\)) & Gaussian Forest (F1\(\uparrow\)) \\
%& Forest (F1\(\uparrow\)) & Forest (F1) & Forest (F1) \\
\hline

PredRNN \citep{wang2017predrnn} & 0.8473$\pm$0.0710 & 0.7369$\pm$0.1102 & 0.7968$\pm$0.1448 \\

PredRNN++ \citep{wang2018predrnn++} & 0.8584$\pm$0.0692 & 0.7373$\pm$0.1357 & 0.8228$\pm$0.1071 \\

ImaGINator \citep{wang2020imaginator} & 0.5436$\pm$0.2507& 0.3608$\pm$0.2289 & 0.4937$\pm$0.2622 \\

SimVP \citep{gao2022simvp} & 0.8823$\pm$0.0528 & 0.7807$\pm$0.1173 & 0.8777$\pm$0.0673 \\

\textbf{AR-NCA} (Ours) & \textbf{0.9232$\pm$0.0468} & \textbf{0.8648$\pm$0.0882} & \textbf{0.9011$\pm$0.0998} \\ 

\hline
\end{tabular}
\vspace{-2mm}
\caption{Comparison with video prediction networks in deterministic forest fire. Training in Dense Forest with the full data and testing in other configurations (Sparse Forest: uniformly distributed but less trees, Gaussian Forest: dense but non-uniformly distributed trees).}
\label{table_videoprediction}
\vspace{-9mm}
\end{table}

\begin{figure}[t]
\vspace{-2mm}
\centering
\includegraphics[width=0.88\columnwidth]{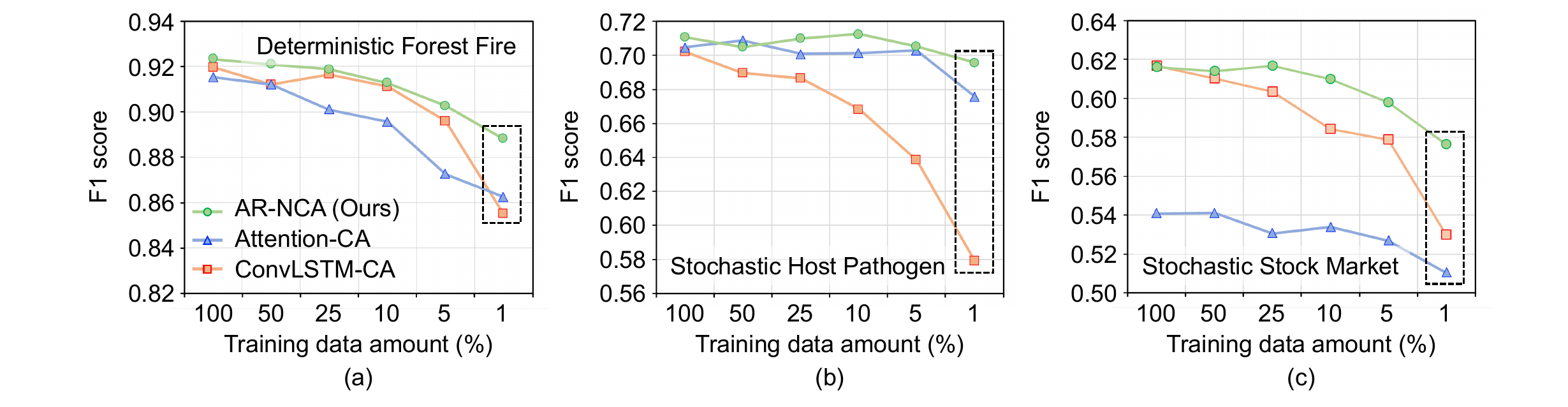}
\vspace*{-5mm}
\caption{Comparison of data efficiency with other NCA networks. The NCA networks are trained and evaluated in the (a) deterministic dense forest with the different amount of training data. Similarly, stochastic (b) host-pathogen and (c) stock market models.}
\label{figure_dataefficiency}
\vspace{-5mm}
\end{figure}
\begin{table}[h]
%\small
\renewcommand{\arraystretch}{1.3}
\setlength{\tabcolsep}{3pt}
\fontsize{8.5pt}{8.5pt}\selectfont
\centering
\begin{tabular}{lcccccc} 

\hline
\multicolumn{1}{l}{Method} & \multicolumn{2}{c}{Stochastic Forest Fire} & \multicolumn{2}{c}{Stochastic Host-Pathogen} & \multicolumn{2}{c}{Stochastic Stock Market}  \\
%(Train$\rightarrow$Test) & & & \\

(Train$\rightarrow$Test Scale) & F1 \(\uparrow\) & AUC (ROC) \(\uparrow\) & F1 \(\uparrow\) & AUC (ROC) \(\uparrow\) & F1 \(\uparrow\) & AUC (ROC) \(\uparrow\) \\ 

\hline
%\multicolumn{7}{c}{100\% Training Data} \\
%\hline

AR-NCA (64$\rightarrow$64) & 0.8987$\pm$0.0598 & 0.9873$\pm$0.0152 & 0.7109$\pm$0.1065 & 0.9578$\pm$0.0308 & 0.6158$\pm$0.1745 & 0.7789$\pm$0.0842 \\ 

AR-NCA (64$\rightarrow$512) & 0.9172$\pm$0.0196 & 0.9878$\pm$0.0079 & 0.7241$\pm$0.0403 & 0.9593$\pm$0.0220 & 0.6177$\pm$0.1467 & 0.7758$\pm$0.0706 \\ 

AR-NCA (48$\rightarrow$512) & 0.9163$\pm$0.0187 & 0.9872$\pm$0.0085 & 0.7196$\pm$0.0427 & 0.9585$\pm$0.0225 & 0.6175$\pm$0.1467 & 0.7758$\pm$0.0705 \\ 

AR-NCA (32$\rightarrow$512) & 0.9114$\pm$0.0215 & 0.9852$\pm$0.0088 & 0.7161$\pm$0.0422 & 0.9579$\pm$0.0227 & 0.6112$\pm$0.1501 & 0.7756$\pm$0.0705 \\ 

\hline
\end{tabular}
\vspace{-4mm}
\caption{Prediction in large systems with AR-NCA  networks trained in small systems in stochastic environments. 64$\rightarrow$64 and 64$\rightarrow$512 indicate the dimension of systems for training (full data) and testing. The table element indicates mean and standard deviation.}
\label{table_scalability}
\vspace{-7.5mm}
\end{table}

\vspace*{-1.5mm}
\paragraph{Comparison with other NCA networks} The question is why AR-NCA outperforms the video prediction networks in extremely data-limited scenarios. It might be due to the cell-based processing of NCAs that effectively captures local interactions. To further understand the properties of AR-NCA and NCAs, we design two other NCA networks, ConvLSTM-CA and Attention-CA. As the prevalent form of NCAs is CNN, we design a ConvLSTM model to process 3$\times$3 cells and preserve the spatial dimension throughout the latent space, named ConvLSTM-CA \citep{shi2015convolutional}. It consists of two convolution layers with a 3$\times$3 kernel and 1$\times$1 kernel for encoding and decoding and one ConvLSTM cell between them with a 3$\times$3 kernel. Recently, attention-based NCA has been proposed \citep{tesfaldet2022attention}. To compare it with our model and also to validate the efficacy of LSTM in AR-NCA, we design another attention-based model by removing LSTM from AR-NCA, named Attention-CA. It is fundamentally analogous to the previous work.

\begin{figure}
\centering
\vspace{-3mm}
\includegraphics[width=\columnwidth]{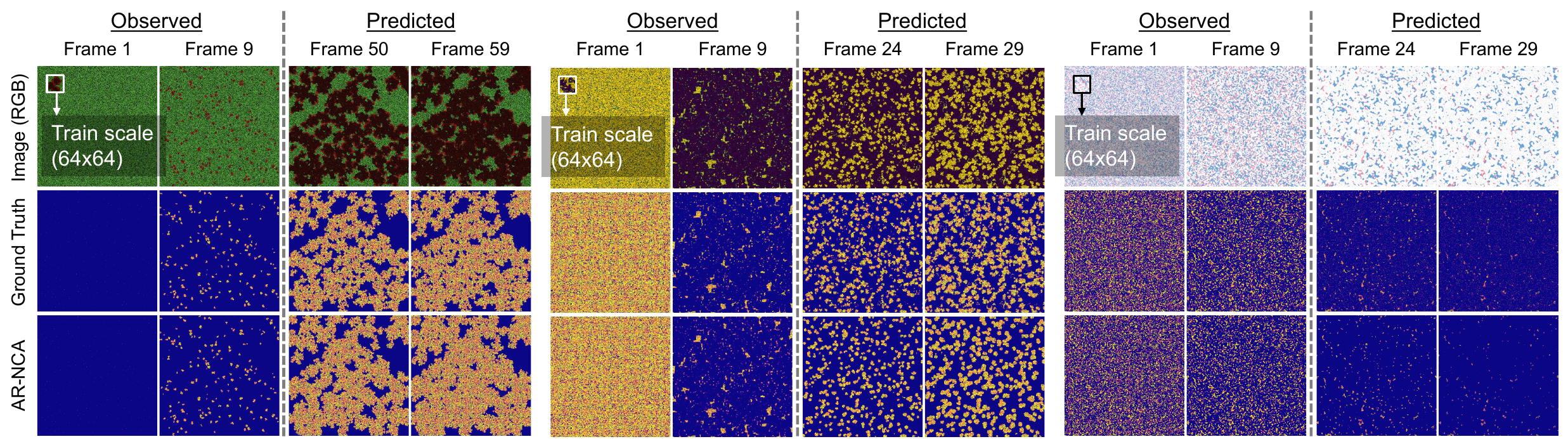}
\vspace{-9mm}
\caption{Visualized prediction results of AR-NCA in stochastic forest fire, host-pathogen, and stock market models from left to right. Training in 64$\times$64 scales (1\% data) and evaluation in 512$\times$512 scales.}
\label{figure_512size}
\end{figure}

\vspace*{-1.5mm}
\paragraph{Data Efficient Interaction Learning} We first explore the data efficiency of the NCA networks in the deterministic forest fire model. We observe that if the full amount of training data is given, the performance of the three networks are similar (Figure \ref{figure_dataefficiency}(a)). We also perform the similar experiments in the stochastic systems, host-pathogen and stock market (Figure \ref{figure_dataefficiency}(b) and (c))\footnote{Empirical results on other stochasticity levels are available in the supplementary material.}. Interestingly, ConvLSTM-CA and Attention-CA also perform better than the video prediction networks with 1\% training data (7 chunks) in the stochastic host-pathogen and stock market. However, there is noticeable performance disparity between AR-NCA and the others, particularly large in the stochastic environments at 1\% training data.

These results demonstrate that the combination of memory module and self-attention is superior than solely using either of them, such as self-attention without memory (Attention-CA) and memory without self-attention (ConvLSTM-CA). It is important to note that both the attention-based networks are slowly degraded with the limited training data than ConvLSTM-CA (Figure \ref{figure_dataefficiency}). This implies that the permutation invariant learning allows the data-efficient discovery of local state transition rules. However, Attention-CA generally exhibits lower accuracy due to the lack of memory function in the network. For example, Attention-CA performs better than ConvLSTM-CA only in the host-pathogen model, where the state transition is not necessarily depending on the past states. In summary, the better performance of AR-NCA can be resulted from two factors; memory function relieves the bottleneck of learning the dynamic interaction in Attention-CA, and permutation invariance promotes data-efficient and well-balanced interaction learning than ConvLSTM-CA.

\vspace*{-1.5mm}
\paragraph{Spatially Scalable Prediction} As long as state transition rules (or interaction) are same, the prediction quality should be preserved even if the spatial scale of  the systems significantly alters (\textit{i.e.}, scalability). We perform comprehensive experiments for the scalability in the three training scales of 32$\times$32, 48$\times$48, and 64$\times$64 and the large test scale of 512$\times$512. Table \ref{table_scalability} showcases the F1 score and AUC across the three stochastic environments with the full amount of training data. The accuracy is not significantly different in the three scales, indicating that AR-NCA successfully learns state transition rules even in the 32$\times$32 scales. Figure \ref{figure_512size} displays the prediction results of AR-NCA in the three systems. The scalable prediction is particularly advantageous to efficient training. For example, training in small systems and then applying to large systems can significantly reduce the training time and memory. Finally, we compare AR-NCA with the other NCAs considering both the data efficiency and scalability. Table \ref{table_comparison_scalability} shows that AR-NCA still outperforms the other NCAs in the three large stochastic environments.

\begin{table}[t]
%\small
\renewcommand{\arraystretch}{1.3}
\setlength{\tabcolsep}{3pt}
\fontsize{8.5pt}{8.5pt}\selectfont
\centering
\begin{tabular}{lcccccc} 

\hline
\multicolumn{1}{l}{Method} & \multicolumn{2}{c}{Stochastic Forest Fire} & \multicolumn{2}{c}{Stochastic Host-Pathogen} & \multicolumn{2}{c}{Stochastic Stock Market}  \\

& F1 \(\uparrow\) & AUC (ROC) \(\uparrow\) & F1 \(\uparrow\) & AUC (ROC) \(\uparrow\) & F1 \(\uparrow\) & AUC (ROC) \(\uparrow\) \\ 

\hline
ConvLSTM-CA & 0.8638$\pm$0.0225 & 0.9708$\pm$0.0161 & 0.5935$\pm$0.0672 & 0.8982$\pm$0.0511 & 0.4663$\pm$0.1825 & 0.7545$\pm$0.0759 \\ 

Attention-CA & 0.8588$\pm$0.0269 & 0.9730$\pm$0.0157 & 0.6903$\pm$0.0487 & 0.9438$\pm$0.0249 & 0.5104$\pm$0.1668 & 0.7667$\pm$0.0699 \\ 

AR-NCA (Ours) & \textbf{0.8921$\pm$0.0194} & \textbf{0.9778$\pm$0.0109} & \textbf{0.7061$\pm$0.0446} & \textbf{0.9463$\pm$0.0275} & \textbf{0.5775$\pm$0.1633} & \textbf{0.7728$\pm$0.0714} \\ 

\hline
\end{tabular}
\vspace{-2mm}
\caption{Comparison with the other NCAs in large stochastic environments. Training in 64$\times$64 scales (1\% data) and evaluation in 512$\times$512 scales.}
\label{table_comparison_scalability}
\vspace{-5mm}
\end{table}

\begin{table}[t]
\renewcommand{\arraystretch}{1.3}
\setlength{\tabcolsep}{3pt}
\fontsize{8.5pt}{8.5pt}\selectfont
\centering
\begin{tabular}{lccccccc} 
\hline
\multicolumn{1}{l}{Method} & \multicolumn{2}{c}{Stochastic Forest Fire} & \multicolumn{2}{c}{Stochastic Host-Pathogen} & \multicolumn{2}{c}{Stochastic Stock Market}  \\

(Embedding Dim.) & F1 \(\uparrow\) & AUC (ROC) \(\uparrow\) & F1 \(\uparrow\)& AUC (ROC) \(\uparrow\)& F1 \(\uparrow\)& AUC (ROC) \(\uparrow\)\\ 

\hline

AR-NCA ($u=16$) & 0.8335$\pm$0.0934 & 0.9696$\pm$0.029 & 0.6810$\pm$0.1160 & 0.9287$\pm$0.0479 & 0.5560$\pm$0.1909 & 0.7738$\pm$0.0852 \\ 

AR-NCA ($u=32$)& \textbf{{0.8672$\pm$0.0729}} & \textbf{0.9768$\pm$0.0220} & 0.6957$\pm$0.1078 & 0.9443$\pm$0.0405 & 0.5762$\pm$0.1887 & 0.7751$\pm$0.0855\\ 

AR-NCA ($u=48$) & 0.8636$\pm$0.0729 & 0.9758$\pm$0.0224 & \textbf{0.7001$\pm$0.1087} & \textbf{0.9485$\pm$0.0372} & \textbf{0.5848$\pm$0.1821} & \textbf{0.7756$\pm$0.0852} \\ 

\hline
\end{tabular}
\vspace{-2mm}
\caption{Comparison in AR-NCA with various embedding dimensions ($u$). The number of parameters are 6,185 ($n=16$), 19,065 ($u=32$), and 39,113 ($u=48$). Trained with 1\% data.}
\label{table_encoding dimension}
\vspace*{-9mm}
\end{table}

\begin{wraptable}{r}{6.4cm}
%\small
\renewcommand{\arraystretch}{1.3}
\setlength{\tabcolsep}{3pt}
\fontsize{8.5pt}{8.5pt}\selectfont
\centering

\vspace{-4mm}
\begin{tabular}{lccccccc} 

\hline
\multicolumn{1}{l}{Method} & \multicolumn{2}{c}{Deterministic Forest Fire} \\

(Neighbor Size) & F1 \(\uparrow\) & AUC (ROC) \(\uparrow\) \\ 

\hline
AR-NCA (3$\times$3) & 0.8880$\pm$0.0610 & 0.9766$\pm$0.0219 \\ 

AR-NCA (5$\times$5) & 0.8786$\pm$0.0592 & 0.9787$\pm$0.0195 \\ 

\hline
\end{tabular}
\vspace{-4mm}
\caption{Comparison in AR-NCA with different neighborhood sizes.}
\label{table_neighbor_size}
\vspace{-7mm}
\end{wraptable}

\vspace*{-1.5mm}
\paragraph{Comparison in AR-NCA} We explore other embedding dimensions ($u$) of AR-NCA other than 32 in the stochastic environments. Table \ref{table_encoding dimension} showcases that relatively high accuracy disparity between $u=16$ and $u=32$ than between $u=32$ and $u=48$. It implies that the limited capacity of $u=16$ model is relieved from $u>=32$ models. From this result, we choose $u=32$ as a baseline dimension (compact but not too small capacity). We also investigate other neighbor size of AR-NCA other than 3$\times$3. Table \ref{table_neighbor_size} compares the neighbor sizes of 3$\times$3 and 5$\times$5 in the deterministic dense forest fire, showing 5$\times$5 does not degrade the performance (lower F1 but higher AUC (ROC)). It is important to mention that both the models do not know about interaction scales in the system, but converging to the similar performance. However, AR-NCA is primarily designed to process 3$\times$3 Moore’s neighborhood, which is the most common configuration in NCAs.

\vspace*{-3mm}
\section{Conclusion}
We present \textbf{A}ttentive \textbf{R}ecurrent  \textbf{N}eural \textbf{C}ellular \textbf{A}utomata (AR-NCA) to learn unknown local state transition rules in the discrete dynamical systems. Our experimental results support that coupling the memory function and permutation invariance in NCA allows data-efficient and scalable learning even in the presence of stochastic interactions. We believe this paper provides an useful empirical background in the predictive modeling of the locally interacting systems. Future work for this paper may explore multi-state systems or even continuous-state systems for more practical applications, such as weather prediction, traffic flow analysis, and ecosystem dynamics modeling.

\newpage

\acks{This material is based on work sponsored by the Office of Naval Research under Grant Number N00014-20-1-2432. The views and conclusions contained in this document are those of the authors and should not be interpreted as representing the official policies, either expressed or implied, of the Office of Naval Research or the U.S. Government.}

\bibliography{l4dc2024}

\end{document}